\documentclass[sigconf]{acmart}

\usepackage{booktabs} 
\usepackage{todos}
\usepackage{enumitem}
\usepackage{graphicx}
\usepackage{listings}

\definecolor{dkgreen}{rgb}{0,0.5,0}
\definecolor{dkred}{rgb}{0.5,0,0}
\definecolor{gray}{rgb}{0.5,0.5,0.5}
\setcopyright{rightsretained}

\begin{document}
\title{Debugging Framework Applications: Benefits and Challenges}

\author{Zack Coker}
\affiliation{Carnegie Mellon University}
\email{zfc@cs.cmu.edu}

\author{David Gray Widder}
\affiliation{University of Oregon}
\email{dwidder@uoregon.edu}

\author{Claire Le Goues, Christopher Bogart, Joshua Sunshine}
\affiliation{Carnegie Mellon University}
\email{clegoues, cbogart, sunshine@cs.cmu.edu}

\newcommand{\zfc}[1]
  {{\scriptsize \textbf{\color{ForestGreen} {ZFC says: #1}}}}

\begin{abstract}
  Aspects of frameworks, such as inversion of control and the structure of
  framework applications, require developers to adjust
  their debugging strategies as compared to debugging sequential programs.
  However, the benefits and challenges of framework debugging are
  not fully understood, and gaining this knowledge could provide
  guidance in debugging strategies and framework tool design.
  To gain
  insight into the process developers use to fix problems in framework applications, we
  performed two human studies investigating how developers fix 
  applications that use a framework API incorrectly. These studies
  focused on the Android \lstinline{Fragment} class and the ROS framework.
  We analyzed the results of the studies using a mixed-methods
  approach, consisting of techniques from grounded theory, qualitative content
  analysis, and case studies.  From our analysis, we produced a theory of the
  benefits and challenges of framework debugging. This theory states that
  developers find inversion of control challenging when debugging but find the
  structure of framework applications helpful. This theory could be used to
  guide strategies for debugging framework applications and framework tool
  designs.
\end{abstract}

%



\maketitle
\newcommand{\clg}[1]{\textcolor{purple}{{ [[#1]]}}}
\newcommand{\cab}[1]{\textcolor{blue}{{ [[#1]]}}}

\section{Introduction}
\label{sec:intro}



Software developers often rely on libraries, structured as functions or
classes, to save development time through reuse.  When a library provides
the central driving event loop of a program, it may be better organized as a 
\emph{framework}. Frameworks organize applications
via a mechanism known as \emph{inversion of control}~\cite{Jaspan11}: 
The core logical
component of the framework calls project-specific code of the application, and only when required.  
This mechanism requires applications conform to a
predefined architecture~\cite{Jaspan11} and interact with code through a defined
interface.  
The main benefits of this approach are time saved through reuse
and from consistent application structure --- 
the provided high-level application architecture, file organization, and standard application
control flow.

Our conjecture is that the aspects that differentiate framework programs from
sequential programs that use libraries 
(the heavy 
use of inversion of control, object protocols, and declarative artifacts) should present unique debugging challenges~\cite{Jaspan11}.  
However, how these 
factors influence a developer's debugging process is not well understood.
Improvements in understanding how these factors influence the debugging process 
could lead to improved framework design, along with improved strategies and tools for
debugging.

We performed an exploratory study to understand how frameworks
help and hinder developers during the debugging process.  To
improve our study's
external validity, we 
investigated two frameworks with different use cases: Android, a mobile development framework, and
the Robotic Operating System (ROS), a robotics framework.
We created debugging tasks based on framework \emph{directives},
statements in framework documentation that specify testable assertions about
the framework's application programming interface (API) and
usually contain nontrivial, unexpected information (e.g.,
``[\lstinline{setArguments}] can only be called before the \lstinline{Fragment} is attached
to its \lstinline{Activity}'')~\cite{Dekel09}.  We focused on directives because these
statements present general framework problems,
instead of application specific issues. We collected directives for
these frameworks and then created debugging tasks based on directive
violations (code that contravenes a directive). We then had participants
perform these debugging tasks and recorded their debugging
process. 

We used a mixed-methods approach to guide our study and analysis,
borrowing techniques from case studies~\cite{Yin09}, 
constructivist grounded theory~\cite{Charmaz14}, 
and qualitative content analysis~\cite{schreier12}. We used this approach to
produce
a theory explaining which framework aspects help and hinder
developers in debugging framework misuses. We also investigated how
the presentation of directive violations to developers can affect the debugging
process.

We found that certain aspects of frameworks benefit developers by reducing the
number of mental steps developers need to achieve a goal, while other
aspects of frameworks present challenges (e.g., inversion of control causes
participants to misdiagnose possible method states).
Our key contributions are:

\begin{itemize}[labelwidth=0.7em, labelsep=0.6em, topsep=0ex, itemsep=0ex, 
		parsep=0ex]
		\item  Results from two studies of humans debugging various directive
          violations taken from the Android \lstinline{Fragment} class, and the
          Robotic Operating System (ROS).
		\item An enumeration of the benefits and challenges in debugging
      misuses of framework APIs.
    \item A theory that explains the benefits and challenges in framework 
      debugging.
\end{itemize}

The rest of this paper is organized as follows.
We discuss the methodology behind our human studies and
theory creation in Section~\ref{sec:methodology}.
We present the results of our analysis of our study results in terms of the the
benefits and challenges of framework  
debugging in Section~\ref{sec:theorySupport}. We present how directive
violation consequences affect debugging difficulty in
Section~\ref{sec:difficultyByConsequence}. 
Section~\ref{sec:theory} presents the theory we produced from the study.
Section~\ref{sec:limitations} discusses the study's threats to validity.
Section~\ref{sec:relatedWork} discusses related work.
Section~\ref{sec:conclusion} concludes.

\section{Methodology}
\label{sec:methodology}

We focused on investigating the unique aspects of debugging framework API
misuse as
compared to debugging sequential programs, and used that knowledge to create a
theory of framework debugging. 
We describe the philosophical
basis of our study in Section~\ref{sec:philosophicalBasis}.
Our source of data was human trials, conducted 
with a case study
procedure, a methodical investigation into a
phenomenon where there may be more variables of interest than data
points~\cite{Yin09}. We chose this method of data collection to
observe and analyze the process participants took when addressing framework debugging
problems. To perform
human trials, we created debugging tasks. We describe the methodology
we used to select frameworks and create these tasks
in Section~\ref{sec:studyFrameworks} and
Section~\ref{sec:taskCreation}. This resulted in seven Android tasks
and three ROS tasks.
Once we created the tasks, we collected study participants and 
conducted human trials, described in
Section~\ref{sec:taskMethodology}.  We coded the
tasks using an iterative process for each framework. First, we coded the
interesting actions of the first few  
participants. Then, we defined coding frames of the interesting actions using
qualitative content analysis, a 
technique that condenses verbal or visual data into important
topics~\cite{schreier12}. After coding both case studies, we performed
theoretical sorting to condense 
the coded data and other sources into a cohesive theory~\cite{Charmaz14}.
  
\subsection{Philosophical Basis}
\label{sec:philosophicalBasis}

We created our framework debugging theory via a mixed-method methodology consisting of 
constructivist grounded theory~\cite{Charmaz14}, qualitative content 
analysis~\cite{schreier12}, and case studies~\cite{Yin09}. Our first guiding
principle for our study approach is based in grounded theory: 
the theory created by the investigation is grounded in the data, but
further investigations may be needed to verify the resulting
theory~\cite{Hoda2012Developing}.  The study also has a philosophical basis in
constructivist grounded theory: the researcher influences the results and there may be
multiple correct theories for the same phenomenon due to different
perspectives~\cite{Charmaz14}.
We chose these philosophical bases for two reasons: (1) while an
exploratory study can provide enough insight for theory formation, further
controlled studies need to be conducted to verify any theory created from an
exploratory investigation and (2) while researchers should try to
remove as many biases as possible from an investigation, it is currently
impossible for a researcher to remove all unconscious biases, which may
influence the study results. We also took precautions to minimize the biases that could
arise from an in-depth literature review, such as trying to make the study
results match a similar study's results, as recommended by grounded theory~\cite{StolGrounded2016}.  
Thus, we conducted a
minimal, initial literature review and later conducted a more
in-depth literature review after finishing the trials.

\subsection{Frameworks in the Study}
\label{sec:studyFrameworks}

\begin{figure}[t]
\centering
  \includegraphics[width=0.40\textwidth]{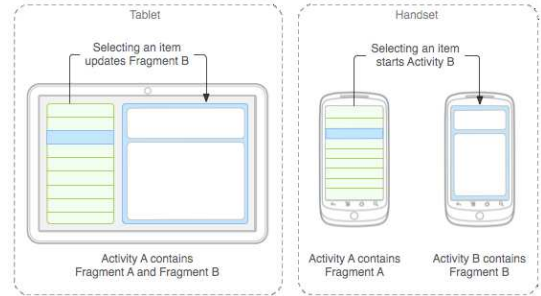}
  \caption[Fragment Caption]{\label{fig:fragmentPic}
An example of the \lstinline{Fragment} class taken from the Android development
  documentation. This diagram demonstrates how the \lstinline{Fragment} class in used in an \lstinline{Activity}.}
\end{figure}

\paragraph{Definitions.}
\emph{Frameworks} provide a set of interfaces and classes that reduce the cost
to achieve a general goal~\cite{Larman}.  Developers create applications to
achieve specific goals by extending frameworks, often by
extending abstract framework methods. The framework typically calls application code 
through \emph{inversion of control}, a design in which
the core framework code, not the application-specific code, controls the data
and execution flow of an application~\cite{Jaspan11}. Frameworks usually achieve
inversion of control through extending abstract methods.  Frameworks
commonly require applications to conform to a specified application
structure. Frameworks also commonly use \emph{object
  protocols}, ordering constraints on calls to an object's
methods~\cite{Beckman11}, and \emph{declarative
  artifacts}, non-source code files that contain configuration
information~\cite{Jaspan11}.

\paragraph{Framework Selection Process.}
We conducted our study using debugging tasks for two frameworks: Android
(version 5.0 Lollipop - API 21, specifically the \lstinline{Fragment} class),
and the Robotic Operating System (ROS) (Kinetic Kame). 

Google's \emph{Android}~\cite{Android} provides a Java framework for developing
mobile applications.  Android is a widely-used, mature framework and has been
released for over seven years.  The Android \lstinline{Fragment} class
represents a reusable component of an Android application's user interface.  A
picture of an Android \lstinline{Fragment} in an example Android application is
shown in Figure~\ref{fig:fragmentPic}, to illustrate its usage.  We began with
the Android framework for three reasons: (1) it is widely-used and
well-established, (2) it makes heavy use of inversion of control and object
protocols, key features that differentiate frameworks from libraries, and (3)
multiple developers express difficulties with the framework, as demonstrated by
searching for StackOverflow\footnote{stackoveflow.com} questions.
In particular, we found that a
large portion of StackOverflow questions focused on the Android
\lstinline{Fragment} class, informing our 
focus on that class in our study.

To improve the generalizability of our claims, we selected ROS as a second
framework for study.  \emph{Robot Operating System}~\cite{ROS} (ROS) is a
framework for creating robotics applications, with a focus on the communication
between various robotic components. ROS applications are built as a collection
of nodes that communicate in an event driven model. ROS is also a mature
framework and has been released for over eight years. ROS is written in both C++
and Python. The ROS framework is both more domain specific and has a
significantly smaller developer base than Android.

Our criteria for the second framework was that it should
focus on a different domain than Android, and have a different
framework architecture.  We decided on the ROS framework for four 
reasons: (1) ROS is designed for robotic applications instead of mobile
applications, 
(2) ROS uses an event based architecture instead of the tiered architecture of 
Android, (3) ROS is written in C++ instead of Java, and (4) ROS has a
smaller user base, but still sufficient users that we could find experienced 
participants for the study.

\subsection{Task Creation Methodology}
\label{sec:taskCreation}
We used violations of framework
directives to create the debugging tasks. Framework directives are testable, non-obvious statement
in a documentation source
about how to use the framework (e.g., ``\lstinline{setHasOptionsMenu(true)} must be 
called to execute an
overridden \lstinline{onCreateOptionsMenu} method''). A directive violation is 
a section of code or application  that does not conform to the testable
statement.
We focused on 
directive violations to improve the chances that our results will generalize,
because framework API misuse errors are not application specific.
Framework directives are also likely to provide situations where participants have
difficulty with a framework, due to the surprising nature of directives.   The
process of extracting directives for both frameworks
consisted of one author extracting directives from the documentation
and another author double-checking the extracted directives, similar to the
coding process in prior work~\cite{StolGrounded2016}.

We first 
collected 45 Android \lstinline{Fragment} directives from three official
documentation sources: (1) the \lstinline{Fragment} page in the developer's
guide,\footnote{developer.android.com/guide/components/fragments.html} (2) the 
\lstinline{Fragment} API
page,\footnote{developer.android.com/reference/android/app/Fragment.html} and (3) the 
\lstinline{Fragment} class's source code. 
11 directives were from the fragment guide, 
19 directives were from the \lstinline{Fragment} API page, and 15 were from the official
\lstinline{Fragment} code. 
We 
extracted directives from documentation statements and error messages in the
code, but only if the directives were testable and described a
non-obvious requirement of the framework.  

To inform our task selection, we investigated
the consequences of
violating Android \lstinline{Fragment} directives and how those
violations were presented to developers.
We created violation scenarios through a multi-step
process: (1) we manually inspected the directives, (2) we created scenarios that
violated the directives, and (3) we manually confirmed the directive violation, 
either with the scenario's output or through
print statements. We then categorized the directives
by the directive violation consequence --- The effect on the
application that the developer would see when that directive was violated.

To select tasks, we searched for StackOverflow questions that cover Android \lstinline{Fragment} directives
from
a wide range of violation consequence categories.
We found seven StackOverflow questions and used the questions to create seven tasks.
The seven tasks were created by taking an Android Lollipop sample 
application\footnote{github.com/googlesamples/android-LNotifications} that
demonstrated the various notifications available in Android Lollipop and
changing the 
application to encompass the scenarios mentioned in the StackOverflow 
questions.

For the ROS framework, we extracted 28 directives from two sources: (1) the official ROS C++ 
documentation,\footnote{wiki.ros.org} and (2) ROS C++ source code.\footnote{docs.ros.org/api}
9 directives were from the documentation and 19 were from the source code.
Due to the relatively low number of online questions about ROS, we were unable
to collect ROS directive scenarios from StackOverflow.  Instead, we choose three
directives that represented materially different cases, and manually created tasks for
each. We created the first and third task by modifying the TurtleSim
scenario,\footnote{http://wiki.ros.org/ROS/Tutorials/UsingRxconsoleRoslaunch} a
two node configuration where a virtual turtle in one window moves and publishes the
movements so the virtual turtle in the other window mimics the movements. The
second task involved a simple, custom-built directory reading application.

In the rest of the paper, the Android tasks and participants will be
prefixed with a ``TA'' and ``PA'' respectively.  The ROS tasks and participants 
will be prefixed with a ``TR'' and ``PR'' respectively.
Table~\ref{table:tasks} lists the number of participants per task and
briefly explains the Android and ROS tasks.

\begin{table*}
\begin{center}
\begin{tabular}{lrlll}
  ~Task & Count & Goal & Violated Directive & Result of Directive Violation \\
\midrule
\rule{0pt}{3ex}
  TA1 & 5 &The participant must connect & Application components must &
  Any attempt to access one of \\
      & & user inputs to the output & have a unique ID to be & the components with the \\ 
      & & message when input components & referenced individually.
      & same ID returned the last \\
      & &initially share the same ID. & & component added.\\
\rule{0pt}{3ex}
  \noindent
  TA2 & 6 &The participant must display the & The application should not   & AndroidStudio displayed a \\
      & & application start time on a & pass time data through & warning and recommend \\
      & & tab without a warning. & the constructor. & a fix.\\ 
\rule{0pt}{3ex}
  TA3 & 4 & The participant must make the & The framework only checks for \
  & The \lstinline$OptionsMenu$ does not appear, \\
      & & framework check for an & an \lstinline$OptionsMenu$ if the application & although
      a \lstinline$OptionsMenu$ is defined. \\
      & & updated \lstinline$OptionsMenu$. & calls
      \lstinline$setHasOptionsMenu(true)$. & \\
\rule{0pt}{3ex}
  TA4 & 5 & The participant must display the & The \lstinline$Fragment$ could only access &  The application crashed \\
   & & application's \lstinline$Activity$ (the entry & the \lstinline$Activity$
   if the \lstinline$Activity$ was & with a notification that the \\
   & & point for an Android application) & attached to the
   \lstinline$Fragment$, and & \lstinline$Activity$ was not set. \\
   & & title in a pop-up message on a& the \lstinline$Activity$ was not
   attached. & \\
   & & specific tab.  & & \\
\rule{0pt}{3ex}
  TA5 & 4 & The participant must fix a problem & A tab's arguments can only & The application crashed, stating \\
  & & that occurs when the application & be set \emph{before} the tab is &
  that arguments can only be\\
  & & tries to change the color of a & accessed. & set before the tab has
  started. \\
  & & button on a tab when the tab& & \\
  & & had been previously accessed.& & \\
\rule{0pt}{3ex}
  TA6 & 3 & The participant must change a & Items should be added to
  the & The \lstinline$OptionsMenu$ would not appear. \\
  & & specified \lstinline$ContextMenu$ to an & \lstinline$OptionsMenu$ in the & \\
  & & \lstinline$OptionsMenu$. & \lstinline$onCreateOptionsMenu$ method.  & \\
\rule{0pt}{3ex}
  TA7 & 5 & The participant must fix an & In the application's current state, & The application crashed with\\
  & & incorrect \lstinline$inflate$ method call. & the last parameter of the \lstinline$inflate$ & a stack trace that pointed \\
  & & & call must be \lstinline$false$. & towards core framework code. \\
\rule{0pt}{3ex}
  TR1 & 8 & The participant must fix an & \lstinline$spinOnce()$
  cannot be used when & A node in the application would \\
  & & incorrect \lstinline$spinOnce()$ call. & the framework should perform &
  quit unexpectedly without an \\
  & & & the callback more than once. & error message. \\
\rule{0pt}{3ex}
  TR2 & 8 & The participant must fix an & Local namespaces are not checked & The
  parameter search returned \\
  & & application node's parameter & if a global namespace is used & that the
  parameter does not exist. \\
  & & access. & in a parameter search. & \\
\rule{0pt}{3ex}
  TR3 & 6 & The participant must fix an & An incorrect message type was & The application crashed with an \\
  & & obsolete message type. & used for this version of ROS. & incorrect type
  declaration error. \\
\rule{0pt}{1ex}
\end{tabular}
\end{center}
\caption{\label{table:tasks} Android and ROS tasks in the human trials. TA
  tasks indicate Android tasks; TR, ROS tasks. ``Count'' shows number of
  participants per task. ``Goal'' indicates task success
  conditions. ``Violated Directive'' is a simplified explanation of
  the violated directive motivating the task. ``Result of Directive Violation''
  explains how the application presented errors to participants.}
\vspace{-0.5cm}
\end{table*}

\subsection{Human Trials Methodology}
\label{sec:taskMethodology}

After obtaining IRB approval, our human trial process started with a 
pre-survey to document
participants' framework experience. We provided participants a Surface
Pro 3 tablet containing the tasks,
and we instructed them to
perform think-aloud debugging, vocalizing what they thought
as they went through the debugging process~\cite{Myers16}.  We assigned a
task to each participant, and asked
them to fix the bug. We did not inform 
participants of the directive violation in the task because we were interested in
also studying the fault localization process.
If participants finished a task and could stay for another 20 minutes,  we asked 
them to attempt another task.  We initially assigned tasks randomly but later
selected tasks that the fewest participants had attempted, to provide a
relatively even task coverage.
For each task, we gave participants the maximum time allowed, but they were able
to quit at any time.
We allowed participants to search online for anything, including the inspiration
for the tasks, but we did not allow them to post questions.
While searching online, no participant found the inspiration for any of the
study's tasks. In addition to asking them to vocalize aloud their thoughts and
strategies during the tasks, we asked participants about their approach in
greater detail at the end of the study.

For the Android study, we collected a convenience sample of 15 participants. 11
of the participants had over 2 years of industrial Java or Android experience,
and 14 of the participants had more than a year of industrial Java
or Android experience.   2 participants were current developers and 13 were
graduate students. For the ROS study, we collected a convenience sample of
12 participants. 9 of the 12 participants preferred the C++ version of ROS
over the Python version.  2 of the participants had more than 2 years
of ROS experience and 5 of the participants had over a year of
experience.  3 of the participants were research staff, 8 of the participants
were graduate students, and 1 was an undergraduate student.

We made several procedure changes between the two case studies. 
For Android, we gave participants time to learn the application before
attempting the tasks, 
while we did not provide a learning period for the ROS tasks.  We
made this change because we found that participants commonly spent the Android
learning period exploring sections of the application that were not relevant to
the tasks. We allowed each Android participant a maximum total study time of
three hours; because the ROS tasks required simple fixes, we set the maximum time in the ROS
sessions to one hour.  In the Android study, we required participants
use the recommended Android Integrated Development
Environment (IDE), AndroidStudio, because it provides warnings for
directive violations.  We did not require participants to use any particular
IDE for the ROS tasks, because ROS does not have a recommended IDE.

\section{Framework Debugging Benefits and Challenges}
\label{sec:theorySupport}

We first present the challenges that developers faced while debugging the tasks:
dynamic challenges
(Section~\ref{sec:dynamicChallenges}), static
challenges (Section~\ref{sec:staticChallenges}), and historical challenges
(Section~\ref{sec:evolutionaryChallenges}). Next, we present the benefits of
framework debugging: dynamic benefits
(Section~\ref{sec:dynamicBenefits}), static benefits
(Section~\ref{sec:staticBenefits}), and historical benefits
(Section~\ref{sec:evolutionaryBenefits}). For each category, we begin 
with a brief
example and then elaborate on interesting cases. Because
the Android and ROS frameworks serve different purposes
(Section~\ref{sec:studyFrameworks}), certain benefits and challenges may have 
only occurred in one framework.  

\subsection{Dynamic Challenges}
\label{sec:dynamicChallenges}

Throughout the study, participants struggled to determine the order in which
a framework executes application code, which increased the difficulty of the
debugging process.  Participants seem to prefer a cause and effect ordering.
However, framework code does not typically follow a sequential ordering, instead
executing application code only when needed. This requires code to be structured
as non-sequential event handlers.  This can create uncertainty about
which parts of project-specific code are called and when, and which project
method will execute ``next.''  Object protocols exacerbate this issue by
requiring participants to understand which states various objects can be in when
the framework calls their code.

\looseness-1
\paragraph{Inversion of Control Issues.} 
In framework programs, application-specific method execution order is not always
transparent to the application developer. This sometimes led participants to misunderstand
an application's control flow, increasing 
the difficulty of the debugging task.
For example, in the ROS study, participants (PR18, PR20) assumed the framework
did not call a section of code, when instead a problem in that code segment
caused the application to terminate earlier than expected. 
In this instance, the framework's inversion of
control led participants to misunderstand the application's behavior, causing
them to waste time while investigating the application.
Inversion of control made it difficult to understand when methods were called and to locate error messages,
and prevented intuitive fixes required modification of framework code.

In both Android and ROS, participants had difficulty understanding the
application control flow. In Android, two participants 
(PA10, PA11) tried to 
use the debugger to understand control flow, but struggled to do so.
Both participants stepped past a current method and were unable
to figure out how to step back into non-framework code.  This led
participant PA10 to reach incorrect conclusions about which code executed in task TA7.
In ROS, while trying to understand how two nodes communicated, PR22 did not
realize that a third node linked two other nodes, because the nodes relied on
the framework to handle communication.  Participant PR22 read four files before
understanding how the framework routed the nodes' communication.
Another participant
(PR23) made incorrect control flow deductions due to the way ROS redirects and filters
statements printed to standard output. Participant PR26 mentioned uncertainty
about how to modify a method because of the states the
application 
could be in when that method was called.

Inversion of control also made localizing errors difficult.
In Android, one participant (PA5) searched for an error message 
thrown by the application, but could not find it in the project.  The search failed
because the error message was 
generated from core framework code, not project code.  

Some problems stemmed from participants' uncertainty about the
hidden ordering of critical framework activity between events.  
When participants (PA4, PA12) saw the \lstinline{getActivity} call 
returned \lstinline{NULL},
they questioned whether the framework had incorrectly constructed its own
reference to the parent \lstinline{Activity}.
In fact, \lstinline{getActivity} was called in 
an event that occurred before the framework had attached the
\lstinline{Activity} to the
\lstinline{Fragment}.

\paragraph{Object Protocols.}
Object protocols are object states that dictate how an object can be used. An 
example of object protocols in Android are lifecycles: state transitions
between starting, active, and stopping for components.
Participants experienced challenges with object protocols (e.g.,
accessing values before they were set). Object protocols are explained and diagrammed explicitly in the 
documentation, but implemented indirectly in the framework code, and invisible 
to non-framework code, which likely led to increased difficulty with object
protocols. Object protocols were more prevalent in the Android
tasks, and thus we observed how object protocols produced framework debugging
challenges in the Android study.

Object protocol issues in tasks TA4 and
TA5 significantly contributed to the amount of time those tasks took (see
Section~\ref{sec:difficultyByConsequence}). 
Most participants assumed the application had
performed an invalid action, rather than an invalid action in a given state.
Object protocol misunderstandings also led participants to incorrectly conclude that certain values were
available for application use.  Three participants (PA4, PA6, PA11) wrote code
to access variables storing
participant-selected times before the participant could have selected those times and
were then confused when the accessed times
did not match the time they selected in testing.   
Participants (PA6, PA10) were confused about the circumstances in which they needed to 
commit and finalize
a \lstinline{Fragment} transaction (as opposed to the cases in which
transactions were automatically committed).

\subsection{Static Challenges}
\label{sec:staticChallenges}

While the static structure of frameworks helped developers in the debugging
process, further discussed in Section~\ref{sec:staticBenefits}, the static structure
also presented multiple challenges to participants.
Participants commonly struggled to understand 
the separation between static structure and dynamic changes,
determine the effects of the
application's static configuration, and use that knowledge to solve problems.
This led to uncertainty about whether errors should be addressed 
by modifying static files, or via a dynamic solution.

Declarative artifacts are non-source code files or the application environment
~\cite{Jaspan11}, such as the XML layout specifications in Android or the XML 
launch file in ROS.  An example of a problem with declarative artifacts is when
participants tried to add a menu using a declarative artifact in a framework that
required menus to be added dynamically. 
Even though there were no errors in the declarative artifacts in
the Android tasks, multiple participants investigated declarative
artifacts to see if they were the source of an error. 

For Android tasks TA3 and TA6, participants created 
\lstinline{OptionsMenu}s. 
Many participants (PA13, PA14, PA15) looked through the Android 
layout editor for
an \lstinline{OptionsMenu} or tried to add an \lstinline{OptionsMenu} to a XML file,
before realizing that it must be added dynamically.  Another
participant (PA9) investigated the \texttt{strings.xml} file after 
an online answer suggested that the problem may lie in an undefined icon title.
Participant PA10 remembered that a specific theme could cause errors
and checked if the
theme caused the error.

In ROS, participants had difficulty understanding how source files map
to executable components, partially because ROS executables
consist of various components (Nodes, Services, and Topics) that do not map
directly to source.
ROS also does not provide an easy or well-known way to find source code 
corresponding to a given component.
%
%
Participant PR16 struggled to understand how the publisher and
subscriber methods in the C++ files integrated with the data
redirections in the launch file.
Other participants (PR17, PR18, PR20, PR22, PR25, PR27) faced
similar difficulties
understanding how the application remapped data among the components.
In one case, Participant PR17 had difficulty finding a source file after
diagnosing the problem: ``I am looking for source code for 
[this node]\ldots~Unfortunately ROS is trying to isolate me from the file system, which I dislike,
because it cannot isolate me fully.'' 16 minutes into the task, Participant PR17 exclaimed ``This
is ridiculous, I can't even find the code that I am supposed to be debugging!''
Participant PR17 eventually used \texttt{grep} to find a node,
more than half an hour into the task.
%

\subsection{Historical Challenges}
\label{sec:evolutionaryChallenges}

Framework changes over time
can increase the difficulty of debugging framework errors. Participants must both
identify gaps in their current understanding and determine
if previous solutions still apply.
For example, participants may find a possible answer for a problem online but may reject
answers that appear out-of-date.  

\paragraph{Legacy Challenges.}
Previous versions of both Android and ROS created issues for several participants.
%
%
In Android, one participant (PA9) questioned whether a feature should be implemented 
in a backwards-compatible way, later discovering that the application was not
configured to work with backwards-compatible components.  A few participants
(PA8, PA9, PA14)
avoided online answers older than two years because
they assumed the answers would no longer apply. 
Other participants (PA1, PA15) mentioned they were familiar with
Android a couple years ago, but there had been many changes to
the framework since they were proficient with it.

Some ROS participants incorrectly diagnosed the obsolete
message type in task TR2 as correct  because they had used it
previously.  Participant PR21 recognized that the message type caused an 
issue, searched
the message type online, and found its official documentation, not realizing
that the documentation was for an older ROS version.  This was a problem because
the documentation indicated that the file was using the
message type correctly.
The participant investigated four other possible error sources before
realizing that a different message type was needed. 

\paragraph{Past Experience.}
While past experience was often helpful, one participant in the Android study (PA10) misdiagnosed 
an error message due to previous experience.  
This caused the participant to conclude to ``not trust your experience.''  

\subsection{Dynamic Benefits}
\label{sec:dynamicBenefits}
Throughout the study, participants commonly used the framework to perform
actions that would have been much more difficult to recreate
without the help of the framework.  When faced with a task, almost all
participants tried to implement the framework method of performing the action
(although PA1, PA3, PA5, PA8, PA11,  implemented custom solutions for certain
tasks, such as implementing a custom message passing solution in TA1). For
example, PA6 in TA1 correctly used the \lstinline{FindFragmentById} method to access user
input, instead of writing code to pipe the data through the application.  This shows 
that developers notice the benefits that framework methods provide in application 
development.

\subsection{Static Benefits}
\label{sec:staticBenefits}

\looseness-1
Study participants found the static organization of the framework helpful
when trying to gain an overview of the application, which helped them
find files of interest more easily than through unstructured search.
In ROS, participants used the launch files as a way to start exploring the
application. For example, participant PR27 looked through the ROS launch files to
understand which nodes are involved in the application.  Participant PR26
mentioned that the participant likes to use launch files to get an overview of
the application.  Multiple participants (PR17, PR18, PR19, PR22, PR26, PR27) used the 
launch files as a way to start the debugging process.  In Android, participants
used the structure of Android application to quickly find resource files and
test case files.  For example, PA8 was able to quickly look up the correct
options menu layout file when writing the required options menu code.
Multiple participants in the Android study (PA1, PA2, PA3, PA6, PA8, PA9, PA10,
PA11, PA13, PA14, PA15) benefited from being able to easily look up application files
to answer questions participants were investigating.

\subsection{Historical Benefits}
\label{sec:evolutionaryBenefits}

Participants often found that past experience was helpful, such as
when they were able to correctly diagnose a ROS 
error simply by looking at the failing section of code.  
Multiple ROS participants (PR17, PR21, PR26, PR27, PR28) were able to diagnose an
error and suggest a working alternative based on past experience. While
working on task TR2, participant PR28 noticed the error in the code and 
said, ``I think the fact that there's a beginning slash means that instead of
looking under this node's namespace it's gonna look under the global namespace
[where] this parameter doesn't exist.'' The participant was correct.
Detailed knowledge of a framework, built up by through experience,
can help mitigate barriers frameworks impose.
Other participants (PR18, PR22, PR26, PR27) stated that past experience shaped
their general ROS debugging strategy. One participant remembered to set framework
environment variables, attributing past environment
problems.
Another participant (PR26) always used \texttt{grep} to find calls to a function
modified over the course of a debugging systems, to guard against unforeseen
side effects, a problem they had faced in the past.

\section{Error Presentation and Debugging Difficulty}
\label{sec:difficultyByConsequence}

We further investigated the relationships between the way errors are presented
to developers and developer debugging success. 
We performed an initial
investigation into the consequences of violating Android \lstinline{Fragment}
directives, where we grouped the consequences into categories. Due to the
limited nature of the investigation, we do not make claims that the categories will
generalize. However, the results are useful as an initial investigation into
the correlation between debugging challenges and error presentation.
We explain the consequence categories (Section~\ref{sec:consequenceCategorization})
and discuss how consequences influenced participant success in the Android and
ROS tasks
(Section~\ref{sec:consequenceDifficulty}).

\subsection{Directive Categorization By Consequence}
\label{sec:consequenceCategorization}

\begin{table}
\begin{center}
\begin{tabular}{lr}
Directive Violation Consequence & Count \\
\midrule
AndroidStudio Warning & 3 \\
Compiler Error & 3 \\
Crash With Reference To Directive & 19 \\
Crash Without Reference To Directive & 2 \\
Expected Action Did Not Occur & 9 \\
No Obvious Effect & 5 \\
Wrong Value Returned & 2 \\
\bottomrule
\end{tabular}
\end{center}
	\caption{\label{table:directiveCategorization}The consequence of
	violating 41 fragment directives. Count is the number of directives in the
  category. One directive violation may produce 
	multiple consequences but each consequence is mutually exclusive.}
\vspace{-0.5cm}
\end{table}

The consequences of violating 41 Android \lstinline{Fragment} directives are shown in
Table~\ref{table:directiveCategorization} and elaborated below.  

\vspace{1ex}
\noindent\textbf{AndroidStudio Warning.}
While the application still compiled with these directive violations,  
if these
directives are violated, AndroidStudio (the recommend Android Integrated
Development Environment) marks the location in code with a
serious warning, and sometimes recommends a possible fix. One example of a
directive in this
category is that classes that subclass the \lstinline{Fragment} class must have a public
no-argument constructor. An application will compile if the class
subclassing \lstinline{Fragment} lacks an empty constructor, but
AndroidStudio displays a warning in the class's source file.

\vspace{1ex}
\noindent\textbf{Compiler Error.}
When these directives were violated, the framework threw a compiler error,
preventing the application from compiling.
This consequence occurred when invalid semantics produced a directive violation.
One example is the case when the documentation 
specified that a method could not be overridden.  The compiler prevented a 
developer from overriding this method because the method was declared 
with the final modifier in the parent class.

\vspace{1ex}
\noindent\textbf{Crash With Reference To Directive.}
When these directives were violated, the application crashed with an 
exception that notified the user of the directive violation either directly
or indirectly. One example of this category is, ``\lstinline{getActivity()} should not be
called when the \lstinline{Fragment} is not attached to the
\lstinline{Activity}''.  If this directive was violated, the application 
crashed with a null return from \lstinline{getActivity()}.
This category contains a high
number of directives because all the directives found in the \lstinline{Fragment} 
class's code were of this type.

\vspace{1ex}
\noindent\textbf{Crash Without Reference To Directive.}
When these directives were violated, the application crashed with an exception
that did not notify the developer that a directive was violated, usually with an
error pointing to where the application crashed instead of the location where
the application needed to be fixed.
Violations in this category occur when a more general exception message is thrown, or 
violating the directive puts the application into an invalid state and the
invalid state is caught in a later line.  One example in this category is when 
the result of the \lstinline{inflate} method is used as the return result for
\lstinline{onCreateView}, the last parameter to the \lstinline{inflate} method
call must be \lstinline{false}.  If this directive was violated, the application
would crash with a stack trace that pointed to internal framework code and not
the \lstinline{inflate} line.

\vspace{1ex}
\noindent\textbf{Expected Action Did Not Occur.}
When these directives were violated, the framework did not execute
the intended effect of the relevant section of
code.  The effect did not occur either because violating the directive caused 
the control flow to
change or the semantics were changed.  For example, one directive states 
that an application will only execute the \lstinline{Fragment}'s
\lstinline{onCreateOptionsMenu} method
if the \lstinline{Fragment} calls \lstinline{hasOptionsMenu(true)} in the
\lstinline{onCreate} method.  If the 
\lstinline{hasOptionsMenu(true)} call is removed, the \lstinline{OptionsMenu} will not appear,
even if the \lstinline{Fragment} overrides
\lstinline{onCreateOptionsMenu}.

\vspace{1ex}
\noindent\textbf{No Obvious Effect}
When these directives were violated, the framework correctly performed the
intended action of the associated code segment without crashing the application.
One example of this category is a directive that states that if a
\lstinline{Fragment} does not have a user interface (UI), then the
\lstinline{Fragment} should be accessed by \lstinline{findFragmentByTag()}, but
the \lstinline{Fragment} without a UI could be accessed by \lstinline{findFragmentById()}
without noticeable consequences.

\vspace{1ex}
\noindent\textbf{Wrong Value Returned.}
When these directives were violated, the application did not crash, but a
reference to a part of the application had been lost or was used incorrectly
Any attempt to use the lost or incorrect reference returned a wrong value.  For
example, when a developer dynamically added a UI element, the developer must
assign a unique tag to the added UI element.  If the added UI element does not
use a unique tag, the new tag overrides the matching tag of a previous the UI
element.  The previous UI element is now unreachable through framework supported
methods.

Violating certain directives can produce multiple consequences 
(e.g., a violation can produce an AndroidStudio
warning and crash with reference to the directive), but each consequence is
mutually exclusive (the same consequence could not be categorized in multiple
categories - an application crash cannot be both classified as crash with
reference to the directive and crash without reference to the directive).  We
found that one directive could be violated in two different ways and produced
three possible consequences.  Two other directives could be violated in two
ways, each with different consequences.

\subsection{Difficulty By Consequence}
\label{sec:consequenceDifficulty}

\begin{table*}
\begin{center}
\begin{tabular}{llrrrl}
  Violation & Time & Sessions & Sessions & Success & \\
  Consequence & (Mean) & Completed & Attempted & Rate (\%)  & Tasks\\
\midrule
  1. Android: Wrong Value Returned & 51 min & 4 & 5 & 80 & TA1\\
  2. Android: Crash With Reference To Directive & 47 min & 3 & 9 & 33 & TA4, TA5\\
  3. Android: Expected Action Did Not Occur & 28 min & 4 & 8  & 50 & TA3, TA6\\
  4. Android: AndroidStudio Warning & 23 min & 6 & 6 & 100 & TA2\\
  5.  Android: Crash Without Reference To Directive & 19 min & 4 & 5 & 80 & TA7\\
        6. ROS: Expected Action Did Not Occur & 49 min & 5 & 8 & 63 & TR1\\
        7. ROS: Wrong Value Returned & 36 min & 5 & 8 & 63 & TR2\\
        8. ROS: Compiler Error & 25 min & 6 & 6 & 100 & TR3\\
\end{tabular}
\end{center}
\caption{\label{table:times} The mean time on task and completion rate of tasks 
    with a given consequence. Time on task includes failed attempts.
	}
\end{table*}

\begin{figure}[t]
\centering
  \includegraphics[width=0.4\textwidth]{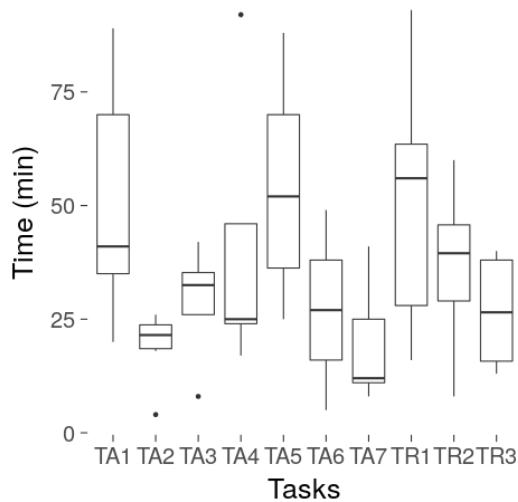}
  \caption{A box-and-whisker plot of the time participants spent on 
  tasks. Time results include failed attempts.}
  \label{fig:taskBoxAndWhisker}
\vspace{-0.5cm}
\end{figure}

We analyzed participant results from both the Android and ROS tasks using the consequence
categories.  Table~\ref{table:times} shows the categorization of each task and
the time spent and participant success rate for each category.
Figure~\ref{fig:taskBoxAndWhisker} shows a box-and-whisker plot of the
participant's time spent on the tasks in the study,
providing further insight on the time range for each task.
We found that there was a significant difference in the mean time to complete
tasks (ranging from 19 to 51 minutes) and the success rate on tasks (ranging from 33\% to
100\%) of difference consequences.

We found that participants struggled to address directive violations that resulted in
a ``wrong value returned'' consequence (tasks TA1 and TR2).
Although four out of five attempts were successful in the Android task, the
attempts took longer than all other consequences (mean: 51 minutes).  
This may also be due to the fact
that participants had to both find a way to uniquely access each value and find
a way to pass them at the appropriate point in the object
protocol. In the case of the ROS task (TR2), where the participant only had to remove an
incorrectly placed slash in a search string, 
the fix was faster but still time consuming (mean: 36 minutes). This task
was a moderately difficult ROS directive violation.


In the Android study, participants had the lowest completion rate when 
the application crashed with reference to the violated directive (tasks TA4 and 
TA5).
For example,
in task TA5, when the participant tried to illegally set the arguments for an
already active fragment, the application crashed with the run time error
``\lstinline{Fragment} Already Active.''
Only three (of nine) participants successfully addressed such
violations, and the mean time taken for these attempts (47 minutes) was much
longer than all other directive violation consequences (with the
exception of the
``wrong value returned'' consequence). 
One reason for this difficulty is that all directives (excluding directives
taken from the
\lstinline{Fragment} source code) that crashed with a reference to the directive 
involved object 
protocols. Although participants were able to find answers on the
object protocol in online questions, these answers did not directly apply to the
task situation.  Instead, participants had to gain a basic understanding of the 
object protocols used in the application before participant knew 
when and how to perform the recommended actions. 

While participants in the Android study were much quicker when an expected
action did not occur (TA3 and TA6), the ROS participants found this violation
consequence to be the most difficult (TR1).
Likely due to the lack of object protocol issues, the Android participants spent
a mean of 28 minutes on these tasks and had a higher success rate
on them (six out of seven attempts successful). 
The ROS participants spent the most time on this task (45 minutes) and tied for
the lowest success rate.  While the fix for TR1 only involved changing one
method call, the participants likely had a more difficult time because
participants had difficulty deducing the fault location from the way the error manifested. In TR1,
the main functionally of the application behaved incorrectly, while only
a single feature of the Android application behaved incorrectly in the Android
tasks.
Participants in the Android study focused on the section of the
application that handled the missing feature, while participants in the ROS study
had to consider the many possible reasons for failure.

The third-fastest (25 minutes) violation consequence was the compiler error task
(task TR3). Participants in this task were the fastest (25 minutes) and were the most 
successful (100\% completion) for the ROS tasks.  
The second-fastest set of attempts (23 minutes) addressed directive violations
that displayed a warning that provided a 
possible solution (task TA2). Participants were able to quickly solve the
problem that caused the warning but still had to spend time implementing the
solution correctly. All 6 attempts at the 
task were successful, tied for the highest completion rate of directive violation 
consequences.
Participants completed task TA7, in which the application crashed without
reference to the directive (task TA7), more quickly than any other task.  As the
error message was largely inscrutable, most participants searched online for the
error after only a brief period investigating the code.  An online search
yielded a quick solution: the task required adding a new parameter to the
inflate method, an easy-to-implement fix based on the online answer.  This
directive violation consequence was fixed the fastest, on average (19 minutes),
and 4 of 5 attempts were successful.

The consequence of violating a directive appears to influence
how long it takes to debug the error as well as how likely a developer is to
succeed in doing so over a short debugging session.
Overall, we observe that
it appears important not only to notify developers of directive violations
but also to help fix them explicitly. 
Often participants knew that certain directives were violated, such as in the
crash with reference to the directive tasks, but they 
did not know how to fix the error. Participants found this frustrating, 
with one participant stating \emph{``Why don't
they tell me the right thing to use?  They tell me it is going to cause a problem
but they don't tell me what the alternative is.''}  

\section{Theory}
\label{sec:theory}
After analyzing the benefits and challenges of frameworks in our study, we condensed those 
aspects into a theory that provides insight
into the framework debugging process. This theory presents the benefits and
challenges of framework debugging in terms of cognitive steps, the number
of mental tasks that a participant must use to achieve a goal.
In this section, we first present the theory created from the human trial
results (Section~\ref{sec:theoryDescription}).  We then discuss our evaluation of
the theory (Section~\ref{sec:theoryEvaluation}).

\subsection{Theory Description}
\label{sec:theoryDescription}

When compared to debugging sequential programs, aspects of the framework
application debugging process reduce the number of cognitive steps required 
to achieve certain goals, and increase the cognitive steps required to achieve
others.
When developers need to find a resource, such as an XML configuration
file, the structure of frameworks
keeps the number of steps required to find the resource to a minimum.  Frameworks can increase the cognitive steps required to debug and fix an error
when participants must understand how to fix an error that has dependencies on
object protocols.  A developer's cognitive load also increases when inversion of
control increases the difficulty in determining the relevant control flows, and
when participants have misconceptions about what the framework is doing outside
application-specific code.

This theory leads to two predictions. The first is that when
debugging framework problems, developers will require fewer steps to do
tasks that involve navigating to files placed in standard framework locations.
Developers will have more difficulty with inversion of
control or object protocol issues.  The second prediction from this
theory is that, when debugging, it will normally be more costly to investigate
framework code to understand the details of how the code works than library code.
This is due to the fact that frameworks are generally
larger than libraries and commonly involve interactions between more components.
This complexity increases the time required to understand portions of a
framework, and increases the chance that participants will investigate an aspect
of the framework that does not apply to the current problem.

\subsection{Theory Evaluation}
\label{sec:theoryEvaluation}

Constructivist grounded theory studies can be evaluated along four
criteria~\cite{Charmaz14}.  The first criterion is \emph{credibility}, which
addresses whether the study has collected enough data to merit its claims.  This
criterion was addressed through our selection of frameworks and tasks, which
cover two diverse frameworks with diverse debugging scenarios.  Having
multiple participants perform each task reduced the risk that our analysis focuses
on anomalous behavior. We attempted to
improve the realism of our scenarios by recreating scenarios from
StackOverflow for the Android tasks.
The second criterion is \emph{originality}, which addresses whether a theory offers
new insights.  While previous studies have investigated what differentiates
framework programming from other types of programming~\cite{Johnson92} and the 
learning issues associated with frameworks~\cite{Ko2004Six}, no previous study 
has investigated how framework aspects affect the debugging process. The
third criterion is \emph{resonance}, or whether the theory makes sense to people in
the associated circumstances. To evaluate this criterion, we contacted six participants
after finalizing the theory and asked them if the theory reflected their
experiences and if the theory provided deeper insights.  All six of the
participants said that the theory reflected their experiences, although one
participant mentioned that the object protocol issues are likely more task
dependent, and not necessarily framework dependent.  Two of the six
participants said that the theory provided deeper insights into the application
debugging process for framework API errors than they had initially. 
The fourth criterion is \emph{usefulness}, which concerns whether the theory provides
interpretations that people can use or build upon. This theory could be useful to
framework debugging tool designers because the theory can help designers focus
on the challenging aspects of framework debugging.  The theory could also be
used to guide novices who are debugging frameworks by focusing them on questions that can
be more easily answered than questions that are more
difficult to answer.

\section{Threats to Validity}
\label{sec:limitations}

\paragraph{Threats to external validity.} We attempt to mitigate the risk that
our theory will fail to 
generalize to other frameworks or languages by
investigating two very different frameworks and a wide range of 
framework debugging problems.  
As stated earlier, our categorization of framework directives by violation consequence 
may not generalize (e.g., other frameworks may not issue formal warnings),
and it may be incomplete; in particular, we did not
consider potential non-functional violation effects, such as degraded performance.

Our constructed tasks may not represent real-world debugging tasks.
This concern was reduced by basing the Android 
tasks on StackOverflow questions. Additionally, participants were new
to the code in each task, possibly leading to unrealistic problems with code familiarity.
We sought to reduce this threat by providing Android 
participants with 
a learning period, but we note that, for example, one participant mentioned that
if the task was encountered in everyday development, it would be preferable
to spend a day reading documentation before
tackling it. As such, time limitations may have influenced our
results.  Finally, the
participants in the study may not represent the population of framework users.  
We attempted to address this limitation by
recruiting participants with experience with the framework: 14 of the
Android study participants had over a year of industrial Android or Java
experience and 7 of the ROS participants had over a year of experience with the
framework.

\paragraph{Threats to internal validity.}
Since this study was exploratory and qualitative, the focus was not on internal
validity. Exploratory studies allow for the 
investigation of a wide array of problems but do not support definitive
cause-effect conclusions.  Participants could freely decide, in a low-risk 
situation, when to quit a task. Participants were also asked to think-aloud, 
and prompted to do so by the researcher. 
These prompts may have altered the route a participant 
would have taken absent the prompt.  Additionally,
the think-aloud aspect of the study may affect how long participants took to
solve the tasks.  We believe that this affected tasks roughly equally, 
such that tasks which took significantly longer than the others are
likely to have taken longer in a non-think-aloud context. Finally, some participants
mentioned they would have been more comfortable if the researcher were not
watching, and if they were able to use their preferred IDE,
operating system, or laptop.  These irritants may have caused
participants to take different routes than they would have in their preferred
environment.

\section{Related Work}
\label{sec:relatedWork}
In this section, we discuss related work in framework investigations, directive
studies, debugging papers, theories of debugging, and grounded theory projects in
software engineering.

\paragraph{Frameworks.}
One of the closest studies to ours investigated the learning barriers
participants face in framework scenarios~\cite{Ko2004Six}, finding some similar
challenges to those we identified, such as coordination barriers, or
difficulties related to using the correct parts of a framework to achieve
framework programming goals.  However, this study focused on general learning
barriers and the relationships among the barriers, instead of focusing on
framework problems.  Our work also covers the benefits of framework debugging
and focuses on framework directive scenarios.  Other prior work has created
formal specifications for framework plugins~\cite{Jaspan09ECOOP} and
investigated the use of declarative artifacts with static
analysis~\cite{Jaspan09RAOOL}.  Another study found that survey respondents
believed they needed to understand the design intent of a framework to use it
effectively~\cite{Robillard09}. Researchers have used
StackOverflow,\footnote{stackoverflow.com} a popular question-and-answer
website, to investigate framework problems~\cite{Wang13}.  Other works have
investigated patterns that appear in framework
development~\cite{Johnson92,Fairbanks06}.  To the best of our knowledge, none of
this prior work specifically addresses debugging.

\paragraph{Directives.}
To the best of our knowledge, prior work on directives has not investigated the
challenges developers face when debugging them. Early work 
investigated how directive knowledge helps developers during coding tasks, and
developed a directive classification scheme based on the topic of the directive
(such as a protocol directive or a performance
directive)~\cite{Dekel09,DekelThesis}. An alternative mechanism for directive
classification focuses on the level of code involved (method,
subclassing, states, etc.)~\cite{Monperrus2012What}. Others have
mined subclassing directives~\cite{Bruch2010Mining}, and fixed 
directives in documentation after analyzing how methods were used in source code\cite{Zhou2017Analyzing}.

\paragraph{Debugging and debugging theories.}
Previous work has found that developers incorporate scent
finding~\cite{Lawrance13} and ask dataflow questions while debugging
generally~\cite{LaToza10}; The ability to easily answer dataflow questions can
significantly reduce the time required in debugging~\cite{Ko2008Debugging}.
Others have found that developers encounter design decisions in the bug fix
process, such as when to fix incorrect data passed between multiple
components~\cite{MurphyHill15}.  Others have explored the debugging of machine
learning programs~\cite{Kulesza10}.  Finally, other researchers have
investigated the challenges of end-user (non-developer) debugging scenarios, and
found that understanding features and testing ideas were important parts of the
process~\cite{Kissinger2006Supporting}.  None of this previous work specifically
focuses on the problems of debugging framework applications.

Prior debugging theories do not capture the unique problems encountered while
debugging frameworks, and thus may only apply to framework debugging at a high
level.  Debugging has been modeled as a trial-and-error process of hypothesis
generation~\cite{Gould75}.  An alternative theory models debugging as a four
stage troubleshooting process: understanding, testing, locating, and fixing;
however this work dealt with programs less than 15 lines long, implying the
``understanding'' stage involved code many orders of magnitude smaller than a
typical modern framework~\cite{Katz87}.

More recent work has characterized debugging as a cyclical process of
gathering and integrating information. One theory describes the process as three
sensemaking loops: the bug-fix sensemaking loop, the environment sensemaking
loop, and the common sense topics and/or domain sensemaking
loop~\cite{Grigoreanu12}.  Another theory models the information gathering
process as searching, collecting, and relating information~\cite{Ko06}.  The
information gathering process of debugging has also been portrayed as various
fact related actions, such as finding and proposing~\cite{LaToza07}.

\paragraph{Grounded Theory in Software Engineering.}
Multiple papers have covered how to present grounded theory papers in software
engineering~\cite{Adolph2008Methodological,Stol2016Grounded,Coleman2007Using,
Hoda2011Grounded}. There are also various software engineering papers that use grounded theory,
such as a theory of the problems developers encounter when moving to a new
software project~\cite{Dagenais2010Moving}, self-organization in an agile
development process~\cite{Hoda2012Developing}, and how communication in an agile
process affects development teams~\cite{Whitworth2007Social}.  None of these
studies have investigated framework debugging, but they provide support for
the use of grounded theory in understanding phenomena in a software engineering context.

\section{Conclusions}
\label{sec:conclusion}

We have presented the results from human trials on framework directive
violation debugging scenarios, and a theory of the benefits and 
challenges of framework debugging. The theory states that certain aspects of the
framework reduce the cognitive steps required by the debugging process, such as 
the static structure of framework applications, which help
developers determine where to find needed files.  The theory also states that
certain aspects of a framework can increase the number of cognitive steps for developers during
debugging, such as the challenges produced by the
inversion of control. In creating this theory, we looked
into the difficulty of solving various directive violations by consequence and
found that assisting developers with directive violations is more complex than 
simplify notifying them of the directive violation in question.
While we believe we have provided sufficient support to justify the creation of a theory,
further work will be necessary to verify it.  This theory provides the
basis for future testable hypotheses, such as investigating framework code often is more
time-consuming than investigating library code while debugging. These
hypotheses can be explored in future studies.

\section{Acknowledgments}
This will be added to a later version of the paper.

\newpage

\bibliographystyle{ACM-Reference-Format}
\bibliography{frameworkstudypaper}

\end{document}